# Nonlinear higher-order polariton topological insulator

YIQI ZHANG,[1] Y. V. KARTASHOV,[2,3] L. TORNER,[2,4] YONGDONG LI,[1] A. FERRANDO[5]

[1]*Key Laboratory for Physical Electronics and Devices of the Ministry of Education & Shaanxi Key Lab of Information Photonic Technique, School of Electronic and Information Engineering, Xi'an Jiaotong University, Xi'an 710049, China*
[2]*ICFO-Institut de Ciències Fotòniques, The Barcelona Institute of Science and Technology, 08860 Castelldefels (Barcelona), Spain*
[3]*Institute of Spectroscopy, Russian Academy of Sciences, Troitsk, Moscow, 108840, Russia*
[4]*Universitat Politècnica de Catalunya, 08034, Barcelona, Spain*
[5]*Departament d'Òptica, Interdisciplinary Modeling Group, InterTech, Universitat de València, 46100 Burjassot (València), Spain*

**Abstract:** We address the resonant response and bistability of the exciton-polariton corner states in a higher-order nonlinear topological insulator realized with kagome arrangement of microcavity pillars. Such states are resonantly excited and exist due to the balance between pump and losses, on the one hand, and between nonlinearity and dispersion in inhomogeneous potential landscape, on the other hand, for pump energy around eigen-energies of corresponding linear localized modes. Localization of the nonlinear corner states in a higher-order topological insulator can be efficiently controlled by tuning pump energy. We link the mechanism of corner state formation with symmetry of the truncated kagome array. Corner states coexist with densely packed edge states, but are well-isolated from them in energy. Nonlinear corner states persist even in the presence of perturbations in corner microcavity pillar.

One of the most representative properties of the conventional topological insulator (TI) is the existence of the link between topological invariants (such as Chern numbers) characterizing spectral bands of bulk TI and the number of the edge states that arise when TI is placed in contact with topologically distinct material, known as a bulk-edge correspondence principle [1]. It was, however, realized that there exist a broad class of higher-order TIs that do not comply with this principle, but nevertheless support topologically protected states [2-9]. Generally, a $d$-dimensional ($d$D) TI supports ($d$D) bulk states and $(d-\ell)$D topological edge states, where $\ell$ has a meaning of the number of dimensions, where edge state is localized due to topology of the system. First-order TIs correspond to $\ell=1$, while for higher-order TIs $\ell>1$. In most cases such higher-order TIs were linear, with the exception in [10,11], where a topological phase is induced only at high amplitudes due to the nonlinearity-dependent intracell hopping rates, not typical for photonic systems.

A powerful platform simultaneously allowing realization of TIs and possessing strong nonlinear effects is provided by polaritons in engineered microcavity structures, such as arrays of microcavity pillars [12-15], where external magnetic field in combination with TE-TM splitting leads to the breakup of the time-reversal symmetry and appearance of the unidirectional edge currents [16-22] that were recently observed experimentally [23]. Edge states in TIs without magnetic fields were observed too [14], but in the 1D arrangement. Polariton condensates are inherently dissipative and require resonant or nonresonant pump for their excitation. The balance between dissipation, pump, nonlinearity, and dispersion leads to a rich variety of dynamical effects and, for the case of the resonant pump, the existence of the pump frequency range where the intracavity intensity becomes bistable [12,20]. Therefore, this system provides an ideal platform for the realization of higher-order polariton TIs and investigation of the impact of nonlinearity on their corner states. Notice that recently coupling between edge and corner states mediated by instabilities was reported in polariton condensates [23].

The goal of this Letter is to study nonlinear effects in higher-order TI realized with kagome arrays of polariton microcavity pillars. A proper shift of the pillars may drive the system into topological phase, where nonlinear corner states can be selectively excited by the resonant pump in the presence of losses and repulsive interactions typical for polaritons. This constitutes first illustration of nonlinear higher-order TI in open, driven system. Strong nonlinearity and resonant pump employed here offer unique advantage of highly selective excitation of the edge or corner nonlinear topological states. Bistability of corner states opens the door for their observation over considerable range of pump frequencies, rather than at isolated frequencies as in linear models.

When polarization effects are neglected, the evolution of polariton condensate in the potential landscape $\mathcal{R}(x,y)$ and under the influence of the resonant pump $\mathcal{H}(t)$ is described by the equation:

$$i\frac{\partial \psi}{\partial t} = -\frac{1}{2}\left(\frac{\partial^2}{\partial x^2}+\frac{\partial^2}{\partial y^2}\right)\psi + |\psi|^2\psi + \mathcal{R} - i\gamma\ \psi + \mathcal{H}. \qquad (1)$$

Here the transverse coordinates $x,y$ are scaled to the characteristic length $L$; the evolution time $t$ is scaled to $\hbar\varepsilon_0^{-1}$, the characteristic energy scale is $\varepsilon_0=\hbar^2/mL^2$; $m$ is the effective polariton mass; $\mathcal{H}=h\exp(-i\epsilon t)$ is the pump term with amplitude $h$; $\epsilon$ is the pump frequency detuning from the bottom of the lower polariton dispersion branch normalized to $\varepsilon_0\hbar^{-1}$; the depth of the potential $\mathcal{R}(x,y)$ is scaled to the characteristic energy $\varepsilon_0$; $\gamma=\hbar/\tau\varepsilon_0$ is the strength of losses, $\tau$ is the polariton lifetime. For the characteristic length of $L=2\,\mu\text{m}$ and

$m = 10^{-34}\,\text{kg}$, one gets the characteristic energy scale of $\varepsilon_0 \sim 0.17\,\text{meV}$, time scale of $\hbar\varepsilon_0^{-1} \sim 3.8\,\text{ps}$, and $\gamma = 0.02$. The potential energy landscape felt by polaritons is represented by the function $\mathcal{R}(x,y) = -\sum_{n,m} \mathcal{Q}(x-x_n, y-y_m)$ describing a kagome micropillar array (see Fig. 1) composed from pillars described by $\mathcal{Q}(x,y) = p\exp\{-[(x-x_m)^2 + (y-y_m)^2]^3/d^6\}$ with $d = 0.5$ ($2\,\mu\text{m}$ diameter) and $p = 10$ ($\sim 1.7\,\text{meV}$ depth of individual potential wells). Kagome array considered here is deformable. The deformation is described by the parameters $d_1$ and $d_2$ defining the shift of each second pillar (see Fig. 2). When $d_1 = d_2 = a/2$ (here $a = 2.5$ that corresponds to $5\,\mu\text{m}$), the structure is not deformed. Nonlinear term in Eq. (1) accounts for the repulsion between polaritons. Our real-world continuous model takes into account all peculiarities of the potential energy landscape in the microcavity, as opposed to treatment based on simplified discrete model.

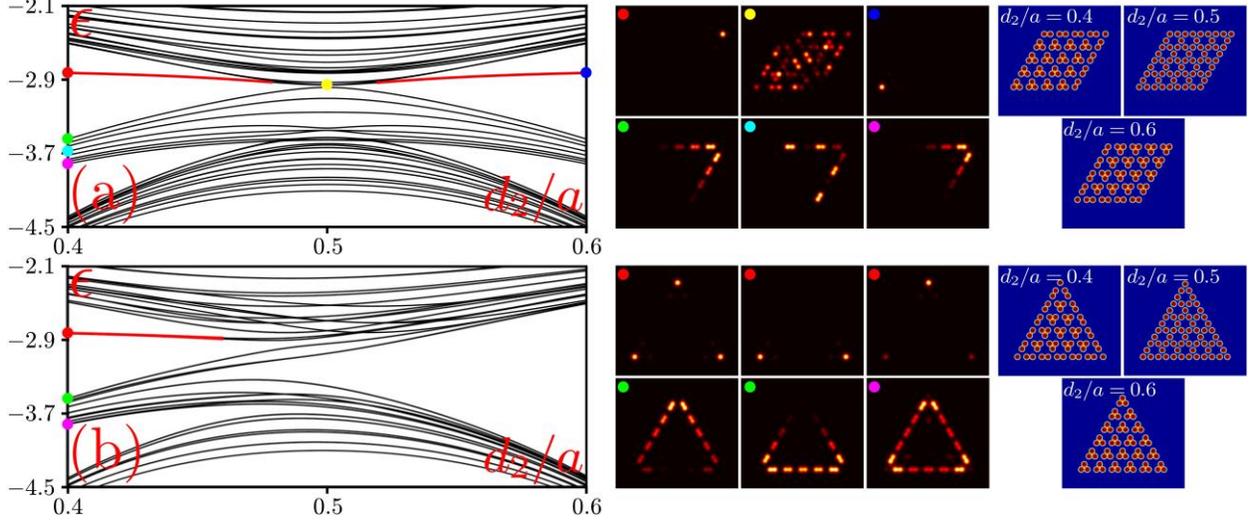

Fig. 1. Band structures (left panels) for rhombic (a) and triangular (b) arrays (two right columns), and representative density distributions (three middle columns) in linear corner and edge states.

To construct higher-order TI we truncate infinite kagome array such, that finite rhombic or triangular configurations form (right column of Fig. 1). The parameters $d_1$ and $d_2$ are used to introduce the topological transition. To illustrate it and to understand which states can be excited by the resonant pump, we consider the spectrum of the *linear modes* without pump and losses $(h, \gamma = 0)$. We find them from Eq. (1) in the form $\psi(x,y,t) = u(x,y)e^{-i\epsilon t}$ solving the resulting linear eigenvalue problem $\epsilon u = -(1/2)(\partial_x^2 + \partial_y^2)u + \mathcal{R}u$. There is a discrete set of modes, whose energies $\epsilon$ depend on $d_2$ in the nontrivial way (Fig 1, left column). Since neighboring pillars start overlapping for large deformations, we vary $d_2$ in the range $[0.4a, 0.6a]$. Band structure for the rhombic array is symmetric about $d_1 = d_2 = a/2$, since rhombic arrays with $d_2 = a/2 \pm \delta$ are mirror images of each other (Fig 1, right column). Topological corner states [red curves in Fig. 1(a)] emerge in the gap and localize in different corners at $d_2 > a/2$ and $d_2 < a/2$. Their localization increases as $|d_2 - a/2|$ grows until they shrink practically to a single pillar [density distributions $|\psi|^2$ for $d_2 = 0.4a$ (red dot) and $d_2 = 0.6a$ (blue dot) are shown in the first row of Fig. 1(a)]. On this reason they are called 0D topological states of this 2D configuration. The middle band in Fig. 1(a) corresponds to the 1D *edge* states that coexist in this geometry with corner ones (but are well-separated from them in energy $\epsilon$ for sufficiently large $|d_2 - a/2|$). They localize near the edges, whose crossing hosts the corner state [see examples for the green, cyan, and magenta dots in the second row of Fig. 1(b)]. The band structure for the triangular array is asymmetric [Fig. 1(b)]. Both corner and edge states appear only when $d_2 < a/2$. Corner states (marked with red color) are now triply degenerate since all three corners are equivalent. All three states corresponding to the red dot in Fig. 1(b) are shown in three middle columns. Edge states from the middle band can be degenerate too and now they occupy all three edges, see examples for the green and magenta dots. Further, we concentrate on the rhombic configuration.

The generation mechanism of the corner states can be understood as a two-step dimensionality reduction process. The first step is the generation of edge states, which appear when the array bandgap occurs at negative $\epsilon$. The continuous field $\psi$ is thus evanescent both outside the rhombus and inside the bulk core. In a discrete version of the model, the amplitude at bulk sites is practically zero. Only at the edge sites the field is non-vanishing, as in Fig. 2. This effectively reduces the dimension of the system from 2D to 1D. The resulting discrete system is a closed rhombic loop that consists of the upper and lower 1D dimer chains, where the AB dimer is the unit cell of the chain, that close the loop by means of a ABA trimer in the right corner (big gray triangle) and with a single B site (small gray triangle) in the left one. The upper and lower 1D dimer chains are Su-Schrieffer-Heeger (SSH) chains with unequal intra- $v$ and inter-cell $w$ coupling constants [24]. Thus, at $d_2 < a/2$ the intracell spacing is larger than the intercell one, $d_1 > d_2$, so that $v < w$.

The 2D finite rhombic array is, in general, asymmetric. Only at $d_1 = d_2 = a/2$ it shows full $C_{2v}$ symmetry: the invariance under $\pi$ rotations ($C_2$) and mirror symmetry with respect to the major and minor rhombus axes ($M_H$ and $M_V$). When $d_2 \neq a/2$, both $C_2$ and $M_V$ are broken and only $M_H$ remains. All modes of rhombic array have to be singlets, since all representations of these symmetry groups are 1D. A 1D loop chain with *odd* number of dimers is asymmetric (Fig. 2). This asymmetry is manifested in the difference between the left connection corner (a monomer) and the right one (a trimer). The upper and lower SSH chains are topologically nontrivial when $v < w$, allowing for the existence of localized edge states for proper chain terminations [24]. Corner states in Fig. 2 can be understood analogously. In the fully dimerized limit $v \to 0$, all dimers and the right trimer are broken because intracell coupling vanishes. Pillar B in the right corner decouples from two chains (green triangle). An isolated corner state with $\epsilon = 0$ localized at this pillar appears. Because $w \neq 0$, states localized on BA pairs of pillars form along the chain with energy $\sim w^{1/2}$. For the same reason, three nonzero energy states at the BAA trimer appear. Consequently, the right corner state is the only allowed zero-energy state separated by a finite gap of width $\sim w^{1/2}$ from the rest of the states. This effective reduction of the dimensionality from 1D to 0D persists also for $v \neq 0$ as long as $v < w$, since the gap remains and the corner state is protected by the topology of the SSH array. The gap decreases as $v$ increases and vanishes when $v = w$ ($d_1 = d_2 = a/2$). This behavior is visible in the calculated band structure of the continuous model at $d_2 < a/2$ [Fig. 1(a)]. The state in the right corner appears in the middle of the gap between two upper bands. Other states (green, cyan, magenta) are modulated extended 1D chain states residing on BA pairs. Clearly, the topology of the system changes for $d_2 > a/2$, when $v > w$, and no corner states are expected to appear at the right corner. However, in this case one can use the specific duality of our structure, associated to the *odd* number of dimers: Mirror reflection of the structure with respect to the $V$ axis and exchange of the roles of $v$ and $w$ couplings leaves the structure invariant. In other words, the structure with $v > w$ is equivalent to the mirror-reflected structure with $v$ and $w$ exchanged. Thus the mirror reflected structure is indeed in a topological phase and it presents a localized state in the *left* corner. This system behaves as a second order TI in 2D with broken reflection symmetry.

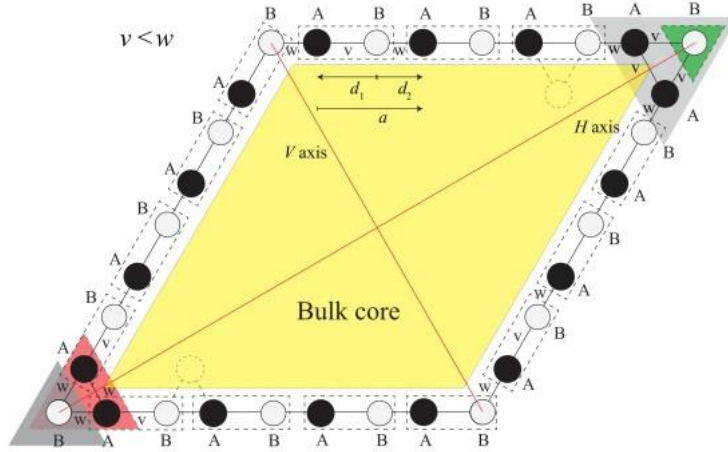

Fig. 2. Schematic representation of the finite rhombic kagome array showing the dimensionally reduced SSH loop chain.

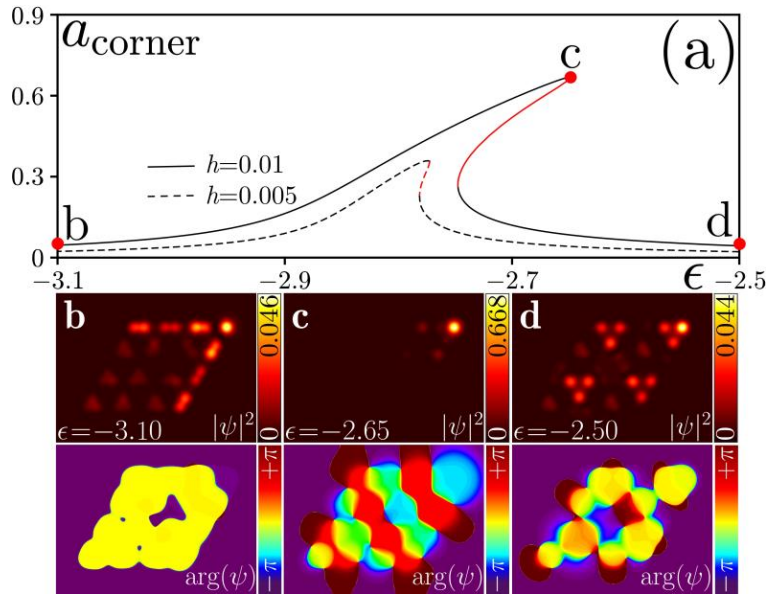

Fig. 3. (a) Amplitude of the nonlinear corner state $a_{\text{corner}}$ at $d_2 = 0.4a$ vs $\epsilon$. Stable branches are shown black, unstable ones are shown red. (b)-(d) Polariton densities and phases corresponding to the red dots in (a) within $x \in [-9,+9]$ and $y \in [-6,+6]$ windows.

The topological properties of the lattice are also characterized by the bulk polarization that can be calculated using the formula $p_j = -S^{-1} \iint_{BZ} A_j d^2k$, where $A_j = -i\langle u | \partial k_j | u \rangle$ is the Berry connection, and $S$ is the area of the first Brillouin zone (BZ). The eigenvectors $u$ can be calculated using the tight-binding Hamiltonian of the system. The system is in topological phase when the polarization components are nonzero and in trivial phase when the polarization is zero [5-8]. In our case $(p'_x, p'_y)$ are calculated in the transformed coordinate system, where BZ is square (see Supplemental Material for topological arguments):

$$(p'_x, p'_y) = \begin{cases} (1/3, 1/3), & \text{for } v<w \ (d_2/a < 0.5) \\ (0,0), & \text{for } v>w \ (d_2/a > 0.5) \end{cases} \quad (2)$$

and agree with the appearance of corner states at $d_2 < a/2$ in Fig. 1(b).

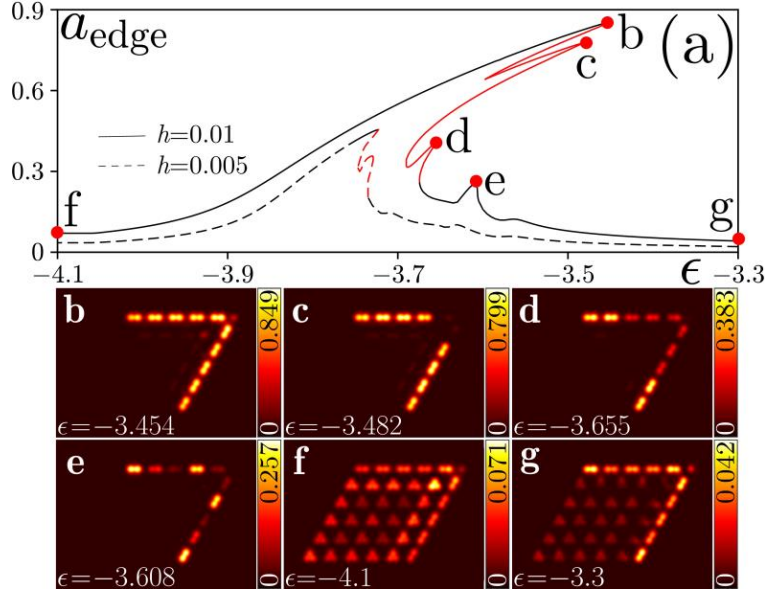

Fig. 4. Resonance curves for nonlinear edge states. Two bottom rows show examples of polariton densities corresponding to the red dots in panel (a) shown within $x \in [-9, +9]$ and $y \in [-6, +6]$ windows. Left and right dashed lines in (a) correspond to the magenta and green dots in Fig. 1(a).

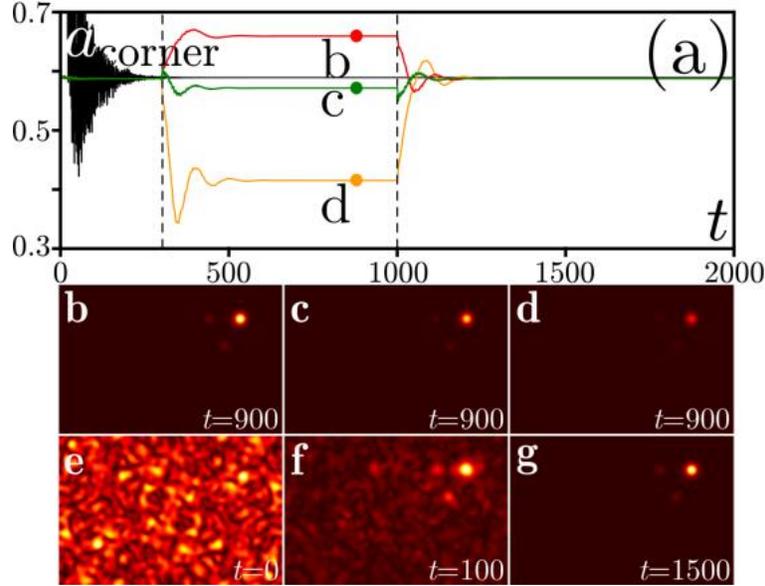

Fig. 5. Robustness of the nonlinear corner state at $\epsilon = -2.7$ from the upper branch of Fig. 3(a) under sudden change of corner pillar at $300 < t < 1000$. Red curve in (a): corner pillar depth is increased by $1\%$, yellow curve: depth is decreased by $2\%$, green curve: corner pillar exhibits $\delta x = 0.05$ shift. (b)-(d) Corresponding density distributions at $t = 900$. Excitation of the corner state by noise [black curve in (a) and density distributions in (e)-(g)].

*Nonlinear* dissipative corner states $\psi = ue^{-i\epsilon t}$ can be excited by the simplest resonant plane-wave pump $\mathcal{H} = he^{-i\epsilon t}$ at zero in-plane momentum $\mathbf{k} = 0$, when polariton-polariton interactions and losses are taken into account. By tuning energy $\epsilon$ one can resonantly excite nonlinear corner or edge states, when $\epsilon$ approaches eigen-energies of the corresponding linear states. The efficiency of the excitation depends on the pump amplitude $h$, and it is maximal for states that are well-separated in energy from other states, i.e. exactly for corner states. Resonant character of the excitation is represented by the dependencies of peak amplitude of the nonlinear state $a_{\text{corner}} = \max|\psi|$ on the pump energy $\epsilon$ [Fig. 3(a)] for the interval of $\epsilon$ values within the gap $[-3.1, -2.5]$, where linear corner states were encountered at $d_2 = 0.4a$.

At small pump amplitudes the resonance occurs exactly at the energy of linear corner state (dashed line). Increasing $h$ leads to a pronounced nonlinearity-induced tilt of resonance and formation of bistable curves, when pump becomes sufficiently strong ($h$ was selected such that the tip of the resonance curve - even for strongest tilt - remains in the topological gap to exclude mixing with bulk modes). Examples of the polariton density and phase distributions in states indicated by the red dots, are shown in Figs. 3(b)-3(d). Far away from the resonance the nonlinear state represents a mixture of several modes and considerably expands into the bulk of the array [Figs. 3(b),(d)], while close to the resonance tip the condensate localizes in the corner pillar [Fig. 3(c)].

Stability of the corner states can be checked by perturbing them with a random noise (up to $10\%$ in amplitude) and modelling evo-lution of such an input up to a long time $t \sim 10^4$, that has allowed us to capture even weak instabilities and to accurately determine stable and unstable branches. Stable and unstable branches are indicated in Fig. 3(a) with black and red colors, respectively. In the bistability regime, the upper and bottom branches are always stable, while the middle branch is unstable. At small pump amplitudes, when bistability is absent, the entire branch of solutions is stable.

In addition to the 0D corner states, one can excite a variety of the 1D edge states by tuning pump energy $\epsilon$ within the band of linear spectrum, where edge states with close energies were encountered [the borders of this band corresponding to the green and magenta dots in Fig. 1(a) are indicated by the dashed lines in Fig. 4(a)]. Because such states have close energies, the pump excites their nonlinear combinations. The resulting resonant curve is very complex, with multiple closely located tips corresponding to the edge states with slightly different density distributions, see Figs. 4(d),(e). At $d_2 = 0.4a$ they all are well-localized near the upper and right edges of the structure. Far from energy band, where linear edge states exist, one excites delocalized nonlinear modes [Figs. 4(f),(g)]. Stable and unstable branches are indicated in Fig. 4(a) by the black and red curves, respectively. One can see that the upper branch of the left outermost resonance is stable, while most of other resonances correspond to unstable states [i.e. the states in Figs. 4(b) and 4(e) are stable, but those in Figs. 4(c) and 4(d) are unstable].

Topological nature of the corner states is confirmed by their robustness with respect to perturbations of the structure (we introduce them by changing the depth and location of the corner pillar). Figure 5 shows that corner state that lies in the topological gap and is therefore topologically protected, just adjusts its amplitude if the depth or location of the corner pillar is changed (for illustrative purposes we introduce this perturbation within time interval $300 < t < 1000$), but is not destroyed unless perturbation is too strong and simply shifts resonant energy value corresponding to localized corner state too far from the pump energy $\epsilon = -2.7$ used in simulations. The amplitude of the corner state increases (decreases) if the depth of the corner pillar increases (decreases), as shown by the red (yellow) curve in Fig. 5. Corner states also withstand small shifts in the corner pillar (green curve in Fig. 5). After perturbation is removed at $t = 1000$, quick recovery to the initial state is observed. Corner states can be also excited from noisy inputs, as shown by the black curve in Fig. 5(a) and by density distributions in Figs. 5(e)-5(f).

Summarizing, we illustrated that topological 0D nonlinear corner states can exist in truncated arrays of polariton microcavity pillars. The mechanism of formation of such states is discussed using symmetry arguments. Resonant pump offers highly efficient and selective excitation of topological corner modes.

**Acknowledgement.** YQZ acknowledges Prof. Ming-Hui Lu, Dr. Hong-Fei Wang and Dr. Yihao Yang for helpful discussions.

**Funding.** NSFC (U1537210); RFBR and DFG project No. 18-502-12080; Spanish MINECO project No. TEC2017-86102-C2-1-R; Fundamental Research Funds for the Central Universities (xzy012019038, xzy022019076); Severo Ochoa Excellence Programme, Fundacio Cellex, Fundacio Mir-Puig, and CERCA.

**Disclosures**. The authors declare no conflicts of interest.

## References

1. T. Ozawa, H. M. Price, A. Amo, N. Goldman, M. Hafezi, L. Lu, M. Rechtsman, D. Schuster, J. Simon, O. Zilberberg, and I. Carusotto, "Topological photonics," Rev. Mod. Phys. **91**, 015006 (2019).
2. C.W. Peterson, W. A. Benalcazar, T. L. Hughes, and G. Bahl, "A quantized microwave quadrupole insulator with topologically protected corner states," Nature **555**, 346-350 (2018).
3. J. Noh, W. A. Benalcazar, S. Huang, M. J. Collins, K. P. Chen, T. L. Hughes, and M. C. Rechtsman, "Topological protection of photonic mid-gap defect modes," Nat. Photon. **12**, 408-415 (2018).
4. F.-F. Li, H.-X. Wang, Z. Xiong, Q. Lou, P. Chen, R.-X. Wu, Y. Poo, J.-H. Jiang, and S. John, "Topological light-trapping on a dislocation," Nat. Commun. **9**, 2462 (2018).
5. A. Hassan, F. Kunst, A. Moritz, G. Andler, E. Bergholtz, and M. Bourennane, "Corner states of light in photonic waveguides," Nat. Photon. **13**, 697-700 (2019).
6. S. Mittal, V. Vikram Orre, G. Zhu, M. A. Gorlach, A. Poddubny, and M. Hafezi, "Photonic quadrupole topological phases," Nat. Photon. **13**, 692-696 (2019).
7. X.-D. Chen, W.-M. Deng, F.-L. Shi, F.-L. Zhao, M. Chen, and J.-W. Dong, "Direct Observation of Corner States in Second-Order Topological Photonic Crystal Slabs," Phys. Rev. Lett. **122**, 233902 (2019).
8. B.-Y. Xie, G.-X. Su, H.-F. Wang, H. Su, X.-P. Shen, P. Zhan, M.-H. Lu, Z.-L. Wang, and Y.-F. Chen, "Visualization of Higher-Order Topological Insulating Phases in Two-Dimensional Dielectric Photonic Crystals," Phys. Rev. Lett. **122**, 233903 (2019).
9. Y. Ota, F. Liu, R. Katsumi, K. Watanabe, K. Wakabayashi, Y. Arakawa, and S. Iwamoto, "Photonic crystal nanocavity based on a topological corner state," Optica **6**, 786-789 (2019).
10. F. Zangeneh-Nejad and R. Fleury, "Nonlinear Second-Order Topological Insulators," Phys. Rev. Lett. **123**, 053902 (2019).
11. G. D'Aguanno, Y. Hadad, D. A. Smirnova, X. Ni, A. B. Khanikaev, and A. Alù, "Nonlinear topological transitions over a metasurface," Phys. Rev. B **100**, 214310 (2019).

12. C. Schneider, K. Winkler, M. D. Fraser, M. Kamp, Y. Yamamoto, E. A. Ostrovskaya, S. Hofling, "Exciton-polariton trapping and potential landscape engineering," Rep. Prog. Phys. **80**, 016503 (2017).
13. V. G. Sala, D. D. Solnyshkov, I. Carusotto, T. Jacqmin, A. Lemaitre, H. Terças, A. Nalitov, M. Abbarchi, E. Galopin, I. Sagnes, J. Bloch, G. Malpuech, and A. Amo, "Spin-orbit coupling for photons and polaritons in microstructures," Phys. Rev. X **5**, 011034 (2015).
14. P. St-Jean, V. Goblot, E. Galopin, A. Lemaître, T. Ozawa, L. Le Gratiet, I. Sagnes, J. Bloch, A. Amo, "Lasing in topological edge states of a 1D lattice," Nat. Photon. **11**, 651–656 (2017).
15. A. V. Nalitov, D. D. Solnyshkov, and G. Malpuech, "Polariton Z Topological Insulator," Phys. Rev. Lett. **114**, 116401 (2015).
16. T. Karzig, C.-E. Bardyn, N. H. Lindner, and G. Refael, "Topological Polaritons," Phys. Rev. X **5**, 031001 (2015).
17. Y. V. Kartashov and D. V. Skryabin, "Modulational instability and solitary waves in polariton topological insulators," Optica **3**, 1228-1236 (2016).
18. O. Bleu, D. D. Solnyshkov, and G. Malpuech, "Interacting quantum fluid in a polariton Chern insulator," Phys. Rev. B **93**, 085438 (2016).
19. D. R. Gulevich, D. Yudin, D. V. Skryabin, I. V. Iorsh, and I. A. Shelykh, "Exploring nonlinear topological states of matter with exciton-polaritons: Edge solitons in kagome lattice," Sci. Rep. **7**, 1780 (2017).
20. Y. V. Kartashov and D. V. Skryabin, "Bistable Topological Insulator with Exciton-Polaritons," Phys. Rev. Lett. **119**, 253904 (2017).
21. Y. V. Kartashov and D. V. Skryabin, "Two-Dimensional Topological Polariton Laser," Phys. Rev. Lett. **122**, 083902 (2019).
22. S. Klembt, T. H. Harder, O. A. Egorov, K. Winkler, R. Ge, M. A. Bandres, M. Emmerling, L. Worschech, T. C. H. Liew, M. Segev, C. Schneider, and S. Höfling, "Exciton-polariton topological insulator," Nature **562**, 552-556 (2018).
23. R. Banerjee, S. Mandal, and T. C. H. Liew, "Coupling between Exciton-Polariton Corner Modes through Edge States," Phys. Rev. Lett. **124**, 063901 (2020).
24. J. K. Asbóth, L. Oroszlány, and A. Pályi. A Short Course on Topological Insulators. (Springer Berlin Heidelberg, New York, NY, 2016).